\documentstyle[12pt,aasms4]{article}
\lefthead{Minezaki et al.}
\righthead{$K$ band Galaxy Counts in SGP}
\begin{document}
\title{$K$-Band Galaxy Counts in the South Galactic Pole Region}
\author{Takeo Minezaki \altaffilmark{1,2,3} and
 Yukiyasu Kobayashi\altaffilmark{1}}
\affil{National Astronomical Observatory,
 Mitaka-shi, Tokyo 181, Japan\\
minezaki@asterope.mtk.nao.ac.jp\\
yuki@merope.mtk.nao.ac.jp}

\author{Yuzuru Yoshii \altaffilmark{1,4}}
\affil{Institute of Astronomy, Faculty of Science, The University of Tokyo,
 Mitaka-shi, Tokyo 181, Japan\\
yoshii@omega.mtk.ioa.s.u-tokyo.ac.jp}

\and

\author{Bruce A. Peterson}
\affil{Mt. Stromlo and Siding Spring Observatories, 
 Institute of Advanced Studies,\\
 The Australian National University, Private Bag, Weston Creek,\\
 A.C.T. 2611, Australia\\
peterson@mso.anu.edu.au}

\altaffiltext{1}{Visiting Astronomer, Siding Spring Observatory operated by
 the Australian National University}
\altaffiltext{2}{Department of Astronomy, 
 School of Science, The University of Tokyo, Bunkyo-ku, Tokyo 113, Japan}
\altaffiltext{3}{current address: Kiso Observatory, Institute of Astronomy,
     Faculty of Science, The University of Tokyo,
     Mitake-mura, Kiso-gun, Nagano 397-01, Japan;\\
     minezaki@kiso.ioa.s.u-tokyo.ac.jp}
\altaffiltext{4}{Research Center for the Early Universe,
 School of Science, The University of Tokyo,
 Bunkyo-ku, Tokyo 113, Japan}

\begin{abstract}

     We present new $K$-band galaxy number counts from $K=13$ to $20.5$
 obtained from $K'$-band surveys in the south galactic pole region,
 which cover 180.8 arcmin$^2$ to a limiting magnitude of $K=19$,
 and 2.21 arcmin$^2$ to $K=21$.
     These are currently the most precise $K$-band galaxy counts
 at $17.5<K<19.0$ because the area of coverage is largest
 among the existing surveys for this magnitude range.
     The completeness and photometry corrections are
 estimated from the recovery of simulated galaxy and stellar profiles
 added to the obtained field image.
 Many simulations were carried out to construct a probability matrix
 which corrects the galaxy counts at the faint-end magnitudes of the surveys
 so the corrected counts can be compared with other observations.
     The $K$-band star counts in the south galactic pole region
 to $K=17.25$ are also presented for use to constrain
 the vertical structure of the Galaxy.

\end{abstract}


\section{Introduction}

    A near-infrared survey of galaxies
 is fundamental for the study of cosmology.
 Merits of using the near-infrared wavelengths are that
 the K corrections of galaxies remain small and nearly independent of
 their morphological type up to $z\sim 2$
 (Cowie et al. 1994; Yoshii \& Peterson 1995),
 and that the evolutionary corrections of galaxies are
 smooth and modeled reliably.
 This is because the luminosity of galaxies in the near-infrared
 is dominated by low mass, late type stars and is less sensitive
 to bursts of star formation (Yoshii \& Takahara 1988),
 and even at large redshift,
 near-infrared observations of galaxies measure their flux
 in well-known optical wavelengths.
 Furthermore, the dust extinction in the near-infrared
 is much smaller than in the optical.

     The $K$-band galaxy number counts have been
 obtained by a number of authors to a variety of depths
 in different areas in order to constrain the geometry of the Universe
 (Gardner, Cowie, \& Wainscoat 1993;
 Cowie et al. 1994; Gardner et al. 1995a, 1995b;
 Glazebrook et al. 1994; McLeod et al. 1995).
 Recently, Djorgovski et al. (1995) and Moustakas et al. (1997)
 surveyed small areas of a few arcmin$^{2}$
 to an extremely faint magnitude of $K\approx 23$ using the KECK telescope, 
 while Gardner et al. (1996) and Huang et al. (1997)
 surveyed very large areas of about 10 degree${}^{2}$
 and presented very precise galaxy counts to $K\approx 16$.

      In this paper, we present new results of $K$-band galaxy counts
 obtained from two surveys in the south galactic pole (SGP) region.
 The bright survey covers 180.8 arcmin${}^{2}$
 to the limiting magnitude of $K=19$,
 and the faint survey covers 2.21 arcmin${}^{2}$ to $K=21$.
      The observations are described in \S2,
 and the image reduction procedures of
 flat fielding, image registration and flux calibration
 are described in \S3.
      The detection and photometry of objects,
 star-galaxy separation, and star counts
 are described in \S4.
      The procedure of correcting the galaxy counts at faint-end magnitudes
 is described in \S5.
      The results of $K$-band galaxy counts are
 presented and discussed in \S6.
      Their cosmological interpretations will be presented in another paper.

\section{Observations}

      The bright and faint surveys were carried out
 during August and September, 1994,
 using the Australian National University's 2.3 m telescope
 at Siding Spring Observatory, Australia,
 equipped with the PICNIC near infrared camera (Kobayashi et al. 1994)
 which was developed at National Astronomical Observatory, Japan.
 PICNIC uses a NICMOS3 array ($256\times 256$ pixels)
 with a field of view of $2.2\times 2.2$ arcmin${}^{2}$ 
 and with a pixel scale of 0.509 arcsec pixel${}^{-1}$.
 In order to reduce the thermal sky background,
 we used a $K'$ filter, which has the same transmission curve
 as the 2MASS $K_{\rm S}$ filter (McLeod et al. 1995).

       The bright survey was centered at $(\alpha,\delta)=
 (0^{\rm h}50^{\rm m}48^{\rm s},-27\arcdeg 43\arcmin 34\arcsec)$ (2000)
 or $(l,b)=(316\fdg27,-89\fdg39)$.
 The observations were made by raster scanning with the telescope. 
 Each scan consisted of eight steps
 with 100 arcsec offsets (30 arcsec overlap) in right ascension
 followed by a step 
 with the same 100 arcsec offsets in declination
 and eight more steps in right ascension in the reverse direction until
 all 64 positions of an eight by eight grid had been observed
 by taking a set of exposures at each
 grid position in the scan.
 Some scans took eight 17.1 s
 exposures at each grid position, others took five, depending
 on the time available to complete the observations.
 The scan pattern was observed 7 times
 to obtain 58 exposures, i.e., 990 s integration per position.

        The faint survey was centered at $(\alpha,\delta)=
 (0^{\rm h}50^{\rm m}54^{\rm s},-27\arcdeg 46\arcmin 42\arcsec)$ (2000)
 or $(l,b)=(313\fdg35,-89\fdg34)$,
 within the area of the bright survey.
 The telescope was shifted randomly in such a pattern
 that no positions were closer than 3 arcsec to each other
 and the positions cover a $30\times 30$ arcsec${}^{2}$ box.
 The integration time of each exposure was also 17.1 s
 and eight exposure were taken at each position.
 The pattern was observed several times so that
 2351 exposures were obtained,
 corresponding to a total integration time of 40000 s.

\section{Reduction}

          The images were reduced using
 IRAF\footnote{
  IRAF is distributed by the NOAO,
  which is operated by the AURA, Inc.,
  under cooperative agreement with the NSF.
 }
 and STSDAS\footnote{
  STSDAS is distributed by the STScI,
  which is operated by the AURA, Inc.,
  for the NASA.
 }.
          The raw $K'$-band images were corrected for thermal scattered light
 by subtracting a sky-image, and corrected for spatial variations
 in the detector response by dividing with a flat field-image after
 subtracting the sky-image. A further correction was applied
 to remove residual thermal stray light and 
 residual sky background variations.

          The construction of the sky-images,
 the flat field-images, and the mosaic required two iterations.
 The objects detected in the mosaiced image constructed for
 the first iteration were masked during
 the construction of the sky-images and flat field-images for
 the second and final iteration.

 The observations for the faint survey and bright survey were
 made in slightly different manners, and this necessitated slightly
 different treatments in the reductions.
 The faint survey observations 
 were made by taking a series of sequential exposures with the telescope
 pointing at essentially the same position in the sky, with
 only small offsets made between every set of eight 17.1 s exposures. 
 The faint survey field
 contained two stars that appeared in each exposure and were 
 used to register all of the individual faint survey images.

 The bright survey observations were made by raster scanning
 with the telescope.
 Each scan consisted of different 64 positions in an eight by eight grid.
 Eight or five in some scans 17.1 s exposures were taken at each grid position.
     The exposures at each grid position from a single scan
 were median combined to create a position-image. If the individual
 exposures contained bright objects, then offsets between the
 different exposures at the same grid position were
 determined from the bright objects and
 used to construct the position-image at that grid position.
 A scan-image was constructed by median combining 
 the 64 position-images of each scan
 after registering them against
 an AAT $I$-band CCD image in the first iteration,
 and against the first iteration 
 $K'$-band mosaiced image in the final iteration.
 In a few cases, a position-image contained no objects suitable 
 for registration, and the relative
 position in the scan was interpolated from the
 registered position-images that preceded and followed in the scan sequence.
 The mosaiced image of the bright survey area was constructed
 by median combining the seven scan-images after
 smoothing a few of the scans obtained in better seeing to
 the characteristic seeing of the survey.  The resultant area of the
 bright survey, after discarding the under-exposed edges,
 is 180.8 arcmin$^2$ and the FWHM of the PSF is 1.5 arcsec. 
 The mosaiced image is shown in Figure {\ref {fig1}} (Plate ?).
 The images formed by median combining sequential blocks of
 100 registered exposures in the faint survey
 were average combined to construct the image of the faint survey area.
 After discarding the under-exposed edges, the
 resultant area of the faint survey is 2.21 arcmin$^2$ and the FWHM of the PSF
 is 1.4 arcsec. The image is shown in Figure {\ref {fig2}} (Plate ?).

 The time and spatial
 variation of the background in the raw $K'$-band exposures does
 not represent simply the variation in detector sensitivity,
 but includes contributions from scattered light, emission
 from dust particles on the optical surfaces, and ambient
 thermal emission from the telescope structure. 
 In order to minimize the influence of
 the time variation of the background, a sky-image was
 subtracted from each exposure.
 For the faint survey, the sky-images were constructed from the
 median combination of blocks of 100 exposures, and subtracted
 from each of the 100 exposures making up the block. 
 For the bright survey, the sky-images were constructed from the
 median combination of exposures taken at the 4 preceding and 4
 following grid positions in the same scan.
 In the final
 iteration, detected objects were masked and the area they covered
 was ignored in constructing the median combination.

 Variations in detector sensitivity were corrected by dividing
 with a flat field-image after subtracting a sky-image from each exposure.
 The flat field-image was constructed from a combination of a dome flat
 and an illumination correction. A dome flat was obtained
 for each night by
 differencing  observations of a white screen with the calibration
 lamp on and off. The dome flat is free of contamination from
 ambient thermal emission, but suffers from uneven illumination.
 An illumination correction was constructed from sky-images and
 dark exposures. The sky-images were corrected for thermal stray
 light, which changed along the rows,
 by subtracting a quadratic function of row number. A dark exposure
 was made with a cold, opaque shutter in the filter wheel
 blocking all external radiation, and was subtracted from
 a stray light corrected sky-image to make a sky flat.
 The flat field-image was then obtained by multiplying the dome flat by
 a sky flat that had been
 divided by the dome flat and smoothed with a $32\times32$ pixel
 box median filter.
 A flat field-image was constructed for each row in each scan of the
 bright survey, and for each block of 100 exposures in 
 the faint survey.
 The effectiveness of the correction for variations in detector
 sensitivity was confirmed by comparing the flux of a standard star
 measured on a 6 by 6 grid of positions on the array.

      The photometric standard stars were taken from
 the UKIRT faint standard stars (Casali \& Hawarden 1992)
 and the IRIS faint standard stars referred to the Carter system
 (Carter \& Meadows 1995).
 The transformation between these two systems is
\begin{equation}
K_{\rm Carter} - K_{\rm UKIRT} = 0.01 - 0.017\times (J-K)
\end{equation}
 (Leggett, Smith, \& Oswalt 1993).
 The colors of observed standard stars were
 $0\lesssim (J-K)\lesssim 0.9$,
 thus the difference of the two systems is
 less than 0.01 mag and negligible.
      Several standard stars were observed at every night,
 and the accuracy of the airmass correction was
 $\sigma _{K}\approx 0.02$ mag.  
 Similarly to McLeod et al. (1995), we detected no color term
 between the $K'$ filter and the $K$ filter
 from the observation of the standard stars.
 We thus estimate the color term
 simply using the filter isophotal wavelengths from Tokunaga (1995)
 (originally from Cohen et al. (1992) for the $H$ and $K$ filter),
 and derived the relation,
\begin{equation}
\Delta _{K}\equiv K'-K=0.04\times (H-K) \;\;\;.
\end{equation}
 Since $H-K=0.2\sim 0.3$ is typical for nearby galaxies
 (Gavazzi \& Trinchieri 1989)
 and $0\lesssim H-K\lesssim 0.8$ is expected for galaxies
 in the bright survey at $z\lesssim 1$ (Eisenhardt \& Lebofsky 1987),
 the magnitude difference between $K'$ and $K$
 was estimated as $\Delta _{K}\lesssim 0.03$.
 The number count error $\Delta _{n}$ propagated from $\Delta _{K}$
 is given by
\begin{equation}
\frac {\Delta _{n}}{n} = 2.3 \alpha \Delta _{K} \;\;\;,
\end{equation}
 where $\alpha \equiv d\log n/dm_{K}$ is the slope of galaxy counts.
 Even if we consider an extremely steep slope of $\alpha = 0.67$
 at $K<16$ from Gardner et al. (1993),
 the difference $\Delta _{K}\lesssim 0.03$ mag yields
 a negligible count error of $\Delta _{n}/n \lesssim 0.05$.
 Therefore we will not distinguish between $K'$ and $K$
 in the remainder of this paper.

\section{Analysis}

\subsection{Detection and Photometry}

     FOCAS (Valdes 1982; Jarvis \& Tyson 1981) was used
 for the detection and the photometry of objects.
 It convolves an image by a user specified filter,
 then collects adjacent pixels which are above
 a user specified threshold from the sky background
 which is determined simultaneously, 
 and decides that a set of connected pixels,
 if more than a user specified number of pixels, is an object.
     These three detection parameters,
 the convolution filter, the surface brightness threshold,
 and the minimum area, were adjusted to maximize the completeness
 while minimizing false detections by simulations.
 We prepared artificial field images
 with a Gaussian random noise field and stellar profiles,
 then tried to detect the artificial objects
 and measure their magnitudes
 using various sets of detection parameters.
 A $\sigma = 1$ pixel Gaussian of $5\times 5$ pixels
 was used for the convolution filter,
 and the area of 5 connected pixels was adopted for the minimum area of object.
 The corresponding surface brightness threshold was
 $\mu _{K}=21.3$ mag arcsec${}^{-2}$ for the bright survey
 and $\mu _{K}=23.3$ mag arcsec${}^{-2}$ for the faint survey.

     FOCAS measures four flux parameters such as core, aperture, isophotal,
 and `total' magnitudes.
 We chose the FOCAS `total' magnitude,
 which measures the flux within a region obtained by
 expanding the detection isophote by a factor of two,
 because it is more stable than other flux parameters.
 This is because the photometric corrections of the aperture magnitude
 become large for nearby bright galaxies,
 and the apertures of the isophotal magnitude for faint objects
 are so small that the large fraction of flux is lost
 with the adopted parameters.

      We carried out many simulations
 to examine the completeness and the photometry.
 We added a small number of artificial objects
 to the resultant mosaiced images
 in order not to change the number density of objects,
 then detected the objects and measured their magnitudes
 with the same FOCAS parameters,
 and compared their FOCAS `total' magnitudes with the input magnitudes
 if they were recovered.
 We repeated this until enough objects were examined.

      For the bright survey,
 both simulated stellar profiles and galaxy profiles
 were considered in the simulations.
      The model PSF was used to construct the simulated stellar profiles,
 and was also used to construct the simulated galaxy profiles
 by convolving the profiles of the model galaxy with it.
      The model PSF was constructed by fitting a moffat function
 to the radial profile of the observational PSF
 which was generated by the average of about twenty bright stars
 in the resultant mosaiced image.
      The model galaxies of different apparent magnitudes were generated
 by placing a galaxy with some absolute magnitude at different redshifts.
    To be consistent with the redshift survey of Songaila et al. (1994),
 which presented $z_{{\rm median}}=0.579\pm 0.1$ at $18<K<19$,
 the absolute magnitude $M_{K}=-23.75$ mag at $z=0$ was determined
 for a model galaxy which should give an apparent magnitude
 of $K=18.5$ at $z=0.6$.
    We adopted $(h, \Omega _{0}, \lambda _{0})=(0.6, 0.2, 0.0)$
 where $h=H_{0}/100$ km s${}^{-1}$ Mpc${}^{-1}$,
 while other choices of the cosmological parameters
 have only small difference at redshifts concerned ($z<1$).
    The K correction was derived by linearly interpolating and extrapolating
 the typical near infrared colors of nearby galaxies,
 $H-K=0.25$ and $J-K=0.95$ (Gavazzi \& Trinchieri 1989),
 thus leading to $-0.56$ mag at $z=0.32$ and $-0.67$ mag at $z=0.79$.
    No evolutionary correction was applied.
    An exponential disk profile was adopted
 for the radial distribution of surface brightness of the model galaxies.
 The internal extinction was neglected
 and the inclination was set randomly.
 The central surface brightness at $z=0$ was adopted as
 $\mu_{K}(0)=17.5$ mag arcsec${}^{-2}$
 (Giovanelli et al. 1995; data originally from 
 Peletier et al. 1994; de Jong \& van der Kruit 1994)
 with an uniform dispersion of $\pm 0.4$ mag.
    The half light radius of the $K=19$ model galaxy
 was 0.75 arcsec and comparable to the PSF.

      For the faint survey, only simulated stellar profiles
 which was constructed from the model PSF were considered in the simulations,
 because galaxies become smaller at fainter magnitude,
 and furthermore the Poisson errors of galaxy counts of the faint survey
 were so large that the difference of estimated completeness
 for detecting either stars or galaxies would be negligible.
 The model PSF was constructed by fitting a moffat function
 to the radial profile of the observational PSF
 which was generated by the average of two bright stars
 in the resultant mosaiced image.

       Five artificial objects consisting of
 either stellar profiles or galaxy profiles
 were added to the resultant image of the bright survey
 for each run of the simulation, and we carried out
 a total of 1760 runs for the stellar profiles
 and 2000 runs for the galaxy profiles at $15.0\le K\le 20.5$.
 Only one artificial object was added to the resultant image of
 the faint survey for each run,
 and we carried out a total of 5000 runs at $18.0\le K\le 22.8$.
       The results of these simulations are presented
 in Figure {\ref {fig3}}.
 The limiting magnitude with 80\% completeness of the bright survey
 was $K=18.8$ for simulated galaxy profiles,
 and $K=19.1$ for simulated stellar profiles.
 This limiting magnitude was $K=21.2$ for the faint survey.

       The number of false detections due to noise
 was estimated by detecting `negative' objects.
 The signs of the resultant images were reversed,
 and then the procedure of detection and photometry was repeated
 with the same FOCAS parameters
 except for the detection threshold below the sky,
 which was adjusted so that the average of the global sky for detection
 was consistent with that for the `positive' detection.
 The estimated false detections in the bright survey
 were very few at the FOCAS `total' magnitude of $K<19$,
 but they started to contribute towards fainter magnitudes.
 Only a few false detections were estimated in the faint survey.

\subsection{Star-Galaxy Separation and Star Counts}

      Stars and galaxies in the bright survey
 were separated based on two morphological parameters,
 the FWHM and the ir1,
 where the FWHM was measured by Gaussian fitting
 of the radial profile by IRAF imexamine task
 and the ir1 was the intensity weighted first moment radius
 which was measured by FOCAS.
     The bright objects at $K<16$ were easily separated
 based on the FWHM only.
     The objects at $16<K<17.5$ were separated
 on the FWHM --- ir1$\times $FWHM diagram as shown in Figure {\ref {fig4}}.
 Since no clear separation was found for fainter objects,
 we did not attempt to separate stars and galaxies at $K>17.5$.
     In the inset in this figure, we also plot
 the simulated stellar and galaxy profiles at $17.0<K<17.5$
 which were used for the completeness and photometry simulations.
 It clearly shows that simulated stellar profiles are well confined
 in the lower-left side of the boundary
 and separated from simulated galaxy profiles.
 No simulated galaxy profiles was miss-classified as stars,
 and only a few simulated stellar profiles were miss-classified as galaxies.
 The distribution in the diagram of simulated stellar profiles
 seems to be slightly different from that of observed stars,
 however,
 the rate of miss-classification from stars to galaxies
 was estimated as only $\le 4$\% and negligible compared to
 the Poisson errors of star counts and galaxy counts,
 even if the boundary was shifted 0.1 pixel smaller in the FWHM.
 The distribution in the diagram of simulated galaxy profiles
 seemed to be somewhat different from that of observed galaxies.
 However, the simulated galaxy profiles modeled a typical galaxy
 and the overall distribution of their morphological parameters
 was similar to that of the observed galaxies.
 The difference in distribution contributed by compact galaxies
 would just become impressive
 when the boundary region for the star-galaxy separation was closed-up.
 Because we found the separation in the plots of the observed objects
 at $K<17.5$,
 and because the estimated rate of miss-classification from stars to galaxies
 were negligible,
 we did not make any corrections for miss-classification of
 the star-galaxy separation.
     The $K$-band star counts to $K=17.25$ obtained from the bright survey
 are tabulated in Table {\ref {tab05}}.
 Since the brightest star ($K\lesssim 11.7$) in the field center
 of the bright survey was used as a guide for the center,
 it was not used for the later analysis.
     In the faint survey, we did not attempt to separate stars and galaxies
 except for the two obvious bright stars.

      In order to estimate the star counts at $K>17.5$
 and their contribution to the total counts,
 the SKY version 4 (Cohen 1994, 1995),
 which is a refinement of the Galaxy model for the infrared point source sky
 originally developed by Wainscoat et al. (1992),
 was fitted to the star counts at $K<17.5$.
 Two parameters of the model,
 the solar displacement $z_{\sun }$ and the halo to disk ratio
 were determined as 
 $z_{\sun }=16.5(\pm 2.5)$ pc
 and halo:disk$=0.56(\pm 0.03)$ times that of
 the SKY version 1 (Wainscoat et al. 1992)
 by the same procedure as that used by Cohen (1995).
 Both values were consistent with those determined by
 Cohen (1995), 15.5 pc and about 0.45 times the SKY version 1,
 based on far-ultraviolet and mid-infrared source counts.
 The details will be described in another paper.

    The star counts and the fitted model are plotted in Figure {\ref {fig5}}.
 By extrapolating the fitted model to $K>17.5$,
 the contributions of the star counts to the total counts
 were estimated as about 7\% at $K=18$, and 5\% at $K=19$,
 which were comparable to the Poisson errors of galaxy counts of
 the bright survey.
 Therefore we subtracted the predicted star counts from the total counts
 to derive the galaxy counts at $K>17.5$ for the bright survey.
 Compared to the Poisson errors of galaxy counts of the faint survey,
 the contributions of star counts were negligible,
 therefore we did not subtract the predicted star counts
 for the faint survey.

\section{Galaxy Counts}

     The galaxy counts at $K<18$ from the bright survey
 were derived as follows:
 The FOCAS `total' magnitudes of objects were corrected
 to total magnitudes based on the simulations,
 and the number of galaxies in a specific magnitude range was counted,
 then the small incompleteness was corrected based on the simulations.
 The galaxies at $17.5<K<18.0$ were not separated from stars,
 therefore their number was estimated by
 subtracting the predicted star counts from the total counts.

      This standard procedure became unsatisfactory
 at faint-end magnitudes of the bright survey,
 because systematic biases existed close to detection limit.
 Because the scatter of error in photometry
 increases rapidly towards fainter magnitudes,
 and because fainter galaxies are more numerous than brighter galaxies,
 more faint galaxies are counted in the brighter magnitude bin
 than bright galaxies are counted in the fainter magnitude bin,
 then the number counts are altered.
 Furthermore, faint objects are preferentially brightened by noise
 to come into detection.

      In order to avoid the problem,
 the galaxy counts at $K=18.25,\ 18.75,\ 19.25$ from the bright survey
 were derived as follows
 (a similar attempt was made by Smail et al. 1995 in optical counts.):
 We first generated the transfer matrix, $T_{ij}$,
 each element of which gives the fraction of galaxies
 with a total magnitude, $m_{\rm total}=m_{j}$,
 that was detected at the FOCAS `total' magnitude, $m_{\rm FOCAS}=m_{i}$,
 based on the simulations.
 We then generated the probability matrix, $P_{ji}$,
 each element of which gives the probability that
 a galaxy detected at the FOCAS `total' magnitude, $m_{\rm FOCAS}=m_{i}$,
 is a galaxy with the total magnitude, $m_{\rm total}=m_{j}$, as
\begin{equation}
P_{ji}=T_{ij}n_{j}/\sum _{k} T_{ik}n_{k}
\end{equation}
 where $n_{j}$ is the number of galaxies at the total magnitude,
 $m_{\rm total}=m_{j}$.
 The number of stars at a specific FOCAS `total' magnitude,
 $N^{\rm s}_{i}=N^{\rm star}\left(m_{\rm FOCAS}=m_{i}\right)$,
 was estimated by multiplying the number of stars at the total magnitude
 predicted by the star count model,
 $N^{\rm s}_{j}=N^{\rm star}\left(m_{\rm total}=m_{j}\right)$,
 by the transfer matrix for stellar profiles, $T^{\rm s}_{ij}$,
 and summing over total magnitudes as
\begin{equation}
N^{\rm s}_{i}=\sum _{j}T^{\rm s}_{ij}N^{\rm s}_{j} \;\;\;.
\end{equation}
 The number of galaxies at a specific FOCAS `total' magnitude,
 $N^{\rm g}_{i}=N^{\rm galaxy}\left(m_{\rm FOCAS}=m_{i}\right)$,
 was estimated by
 subtracting the estimated number of stars, $N^{\rm s}_{i}$,
 and the number of false detections, $N^{\rm f}_{i}$,
 from the total number of objects, $N^{\rm t}_{i}$,
 at the specific FOCAS `total' magnitude,
 yielding $N^{\rm g}_{i}=N^{\rm t}_{i}-N^{\rm s}_{i}-N^{\rm f}_{i}$.
 Then the number of detected galaxies with a specific total magnitude,
 $N^{\rm g}_{j}=N^{\rm galaxy}\left(m_{\rm total}=m_{j}\right)$,
 was estimated by multiplying
 the number of galaxies at the FOCAS `total' magnitude, $N^{\rm g}_{i}$,
 by the probability matrix for galaxy profiles, $P^{\rm g}_{ji}$,
 and summing over FOCAS `total' magnitudes as
\begin{equation}
N^{\rm g}_{j}=\sum _{i}P^{\rm g}_{ji}N^{\rm g}_{i} \;\;\;.
\end{equation}
 We corrected $N^{\rm g}_{j}$ for the incompleteness
 and finally derived the galaxy count.

     The slope index of galaxy counts, $\alpha $, was presumed
 for the calculation of the probability matrix, $P_{ji}$,
 as $\alpha _{1}=0.67$ at $K<16$, $\alpha _{2}=0.49$ at $16<K<18$,
 and the slope index $\alpha _{3}$ at $K>18$ was left as a free parameter,
 because the matrix $P_{ji}$ at the magnitude concerned was
 dependent almost only on $\alpha _{3}$ and
 independent of $\alpha _{1}$ and $\alpha _{2}$.
     The parameter $\alpha _{3}$ was then adjusted to a value of
 $\alpha _{3}=0.28$ to be consistent with the derived slope
 for which $\alpha =0.276$ from $K=18.25$ to $18.75$ or
 $\alpha =0.277$ from $K=18.25$ to $19.25$.
 By this procedure, the slope index $\alpha _{3}$
 and therefore the galaxy counts were well determined.
 For example, an assumed slope of $\alpha _{3}=0.40$ leads to the derived slope
 for which $\alpha =0.29$ from $K=18.25$ to $18.75$ or
 $\alpha =0.32$ from $K=18.25$ to $19.25$.
     A slope of $\alpha = 0.28$ agrees well with $\alpha =0.26$ at $K>18$
 found by Gardner et al. (1993).

      Following the standard procedures,
 the galaxy counts at $K<20$ from the faint survey were corrected
 for incompleteness and magnitude difference arising from photometry errors,
 and the galaxy counts at $20<K<22$
 were derived using the probability matrix, $P_{ji}$.
 We assumed $\alpha _{3}=0.26$ at $K>18$ (Gardner et al. 1993)
 for the estimation of the $P_{ji}$,
 because the Poisson errors of the galaxy counts at $K>20$ were too large
 to determine a more precise slope of $\alpha _{3}$.
       Since the predicted contribution of star counts was
 negligible compared to the Poisson error of the galaxy counts,
 the total number of objects except for the two stars
 was used as the number of galaxies.

\section{Results}

      The $K$-band galaxy counts we obtained
 are tabulated in Table {\ref {tab1}}.
 The raw counts of galaxies of the bright survey at $K>16.5$
 and those of the faint survey at $K>20.0$ are not integers
 because the predicted star counts were subtracted 
 and the probability matrix, $P_{ji}$, was used.
 The errors given for the counts include only
 the Poisson errors defined as a square root of
 the raw number of objects and false detections.
 The $K$-band galaxy counts at $16<K<22$ are plotted in Figure {\ref {fig6}},
 to be compared with other observations.
 It should be noted that the faintest points of each surveys
 are unreliable because their completenesses is small,
 $\lesssim $ 50\%, and large corrections were needed.

       We estimated the field to field variations of galaxy counts
 due to clustering from angular correlation functions. 
 For an angular correlation function of the power-law form
 $w(\theta )=A_w\theta ^{-\gamma}$
 and a circular top-hat window function of angular radius of $\theta _{0}$,
 the rms variation due to clustering is
\begin{equation}
\sigma _{w}=f(\gamma )w(\theta _{0})^{1/2}\bar {N}
\end{equation}
 where $f(\gamma )\sim 1$ and $\bar {N}$ is the mean number of galaxies.
 From Lidman \& Peterson (1996)
 the angular correlation was $\log _{10}w(\theta _{0})\approx -2.5$
 for $21<I<22$ and $\theta _{0}=7.6$ arcmin,
 each of which corresponds to $18<K<19$ where $I-K\sim 3$ (Gardner 1995b)
 and to the area of the bright survey.
 Then the rms variation due to clustering of the bright survey
 at its faintest magnitudes was estimated as $\sigma _{w}\sim 0.056\bar {N}$,
 which was comparable to the Poisson errors.
 The variation due to clustering of the faint survey was estimated
 from the angular correlation of
 $w(\theta )=(\theta /1\farcs 4 )^{-0.8}$ for $K\le 21.5$
 measured by Carlberg et al. (1997) from the survey area of 27 arcmin$^2$.
 The angular radius corresponding to the area of the faint survey is
 $\theta _{0} = 50$ arcsec and its rms variation due to clustering
 was estimated as $\sigma _{w}\sim 0.24\bar {N}$ or 0.09 dex.
 Thus the galaxy counts of the faint survey could be significantly
 affected by clustering.

       The $K$-band galaxy counts obtained from the bright survey
 are most precise at $17.5<K<19.0$ because of its large survey area,
 and agree well with other observations.
 These counts are therefore important to model
 deeper galaxy counts at $K>20$ in small survey areas
 to constrain the geometry of the Universe.
 In addition, we confirm that the steep increase of
 the galaxy counts shows a turnover around $K\sim 18$,
 as Gardner et al. (1993) found that the slope changed
 from $\alpha = 0.67$ to $\alpha =0.26$ at $K\approx 17$.
 This indicates that the galaxies around $K=17-18$ have the largest
 contribution to the extragalactic background radiation in the $K$-band,
 because at this magnitude the slope of the integrated luminosity
 begins to converge.

       The $K$-band galaxy counts obtained from the faint survey
 have large errors,
 and they are slightly lower than other observations
 and the bright survey as shown in Figure {\ref {fig6}}.
 However, the faint survey is subject to large Poisson errors
 and field to field variations due to clustering as described before
 because of its small survey area.
 In addition, we can see in Figure {\ref {fig1}} that
 objects are more sparsely distributed in the field of the faint survey 
 than in the rest of the bright survey.
    Therefore, when these uncertainties are considered,
 the galaxy counts from the faint survey
 are still consistent with other observations.

       In summary, we present new $K$-band galaxy number counts
 obtained from the $K'$-band surveys in the SGP region.
 The completeness and photometry corrections were estimated with simulations,
 and the galaxy counts were derived
 using the probability matrix, $P_{ji}$,
 at the faint-end magnitudes close to the detection limits
 in order to compensate for the photometry errors.
     The bright survey provides galaxy counts to $K=18.75$,
 and they agree very well with other observations.
     The faint survey provides galaxy counts to $K=20.5$,
 and they are slightly lower compared to other observations
 and to the bright survey.
 However, when all uncertainties are considered,
 they are still consistent with other observations.
     We also present the star counts towards the SGP at $K<17.5$
 obtained from the bright survey, which are important for
 studying the vertical structure of the Galaxy.

\acknowledgments

    We acknowledge M. Cohen for fitting the SKY version 4 model
 to our star counts,
    and the SSO staffs and K. Nakamura
 for their technical support for the observation.
    Y. Y. acknowledges the financial support from
 the Yamada Science Foundation for transport of
 the PICNIC camera and the Japanese IR team to Australia.
    T. M. was supported by the Grant-in Aid for JSPS Fellows
 by the Ministry of Education, Science, and Culture,
    and this work has been supported in part by the Grant-in Aid for
 Center of Excellence Research (07CE2002) of
 the Ministry of Education, Science, and Culture.

\clearpage
\figcaption[sgp1_fig1.ps]{
 The $K'$-band mosaiced image of the bright survey area
 in the south galactic pole (SGP) region.
 The large white box encloses the area of 180.8 arcmin$^{2}$
 in which the galaxy counts were obtained.
 The small box at the lower left encloses the area of the faint survey.
 The overlapped regions between neighboring scan positions
 which have larger integration time
 can be seen as the areas of smaller noise background
 and make a checkered pattern.
 \label{fig1}}

\figcaption[sgp1_fig2.ps]{
 The $K'$-band image of the faint survey area in the SGP region.
 The white box encloses the area of 2.21 arcmin$^{2}$
 in which the galaxy counts were obtained.
 Two brightest objects seen in the area are stars.
 \label{fig2}}

\figcaption[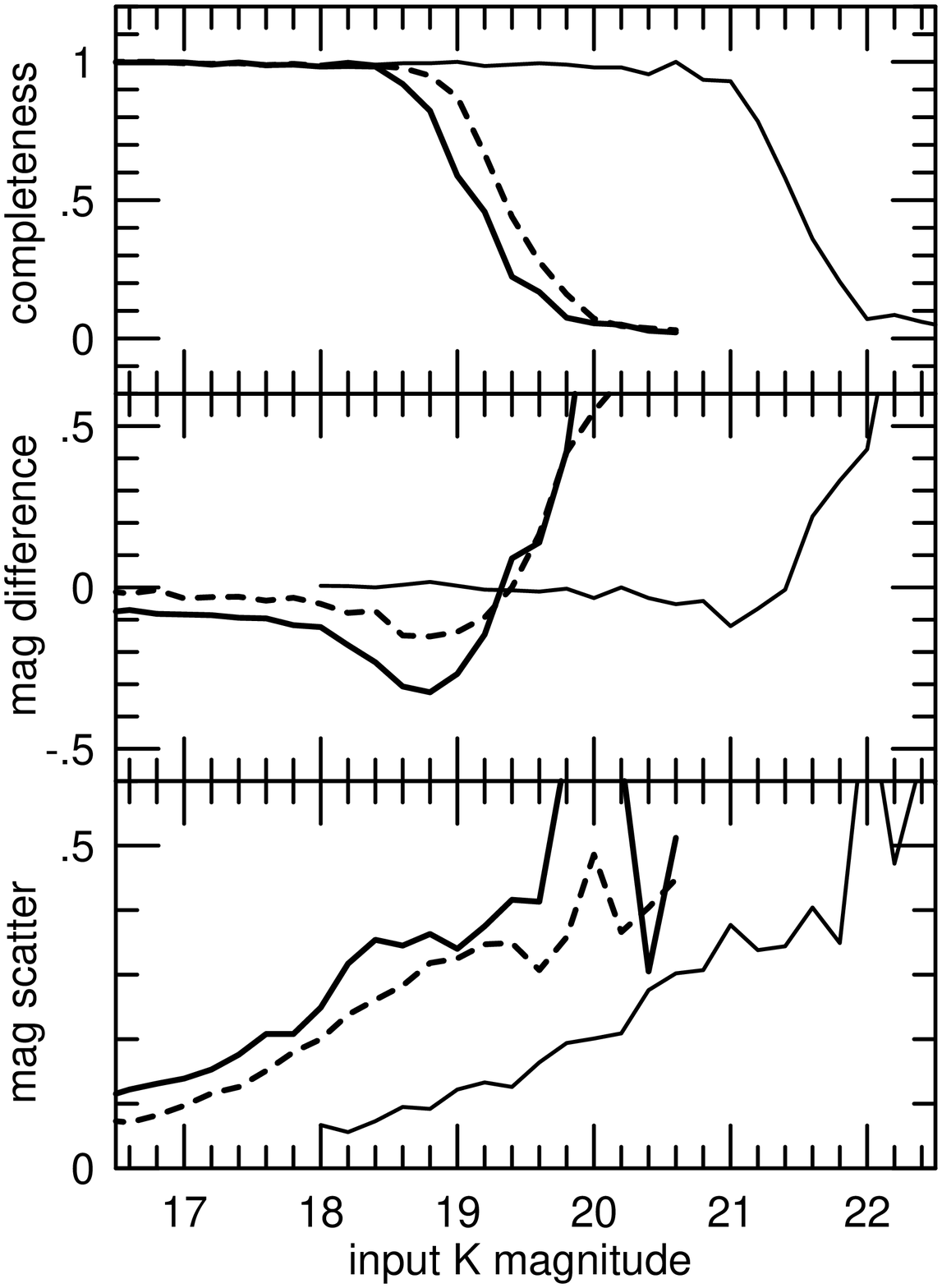]{
 Completeness and photometry estimated from the simulations
 of detecting artificial objects added to the resultant mosaiced images
 of the bright and faint surveys.
 The detection rate or the completeness ({\it top panel}),
 the average of magnitude difference between
 the input magnitude minus the FOCAS `total' magnitude
 of detected objects ({\it middle panel}), and
 the magnitude scatter or the root-mean-square of magnitude differences
 ({\it bottom panel})
 are shown as a function of the input magnitude of objects.
 The dashed and solid lines show the results for the bright survey
 for artificial stellar profiles and artificial galaxy profiles, respectively.
 The thin solid line shows the result for the faint survey
 with the only use of artificial stellar profiles.
 \label{fig3}}

\figcaption[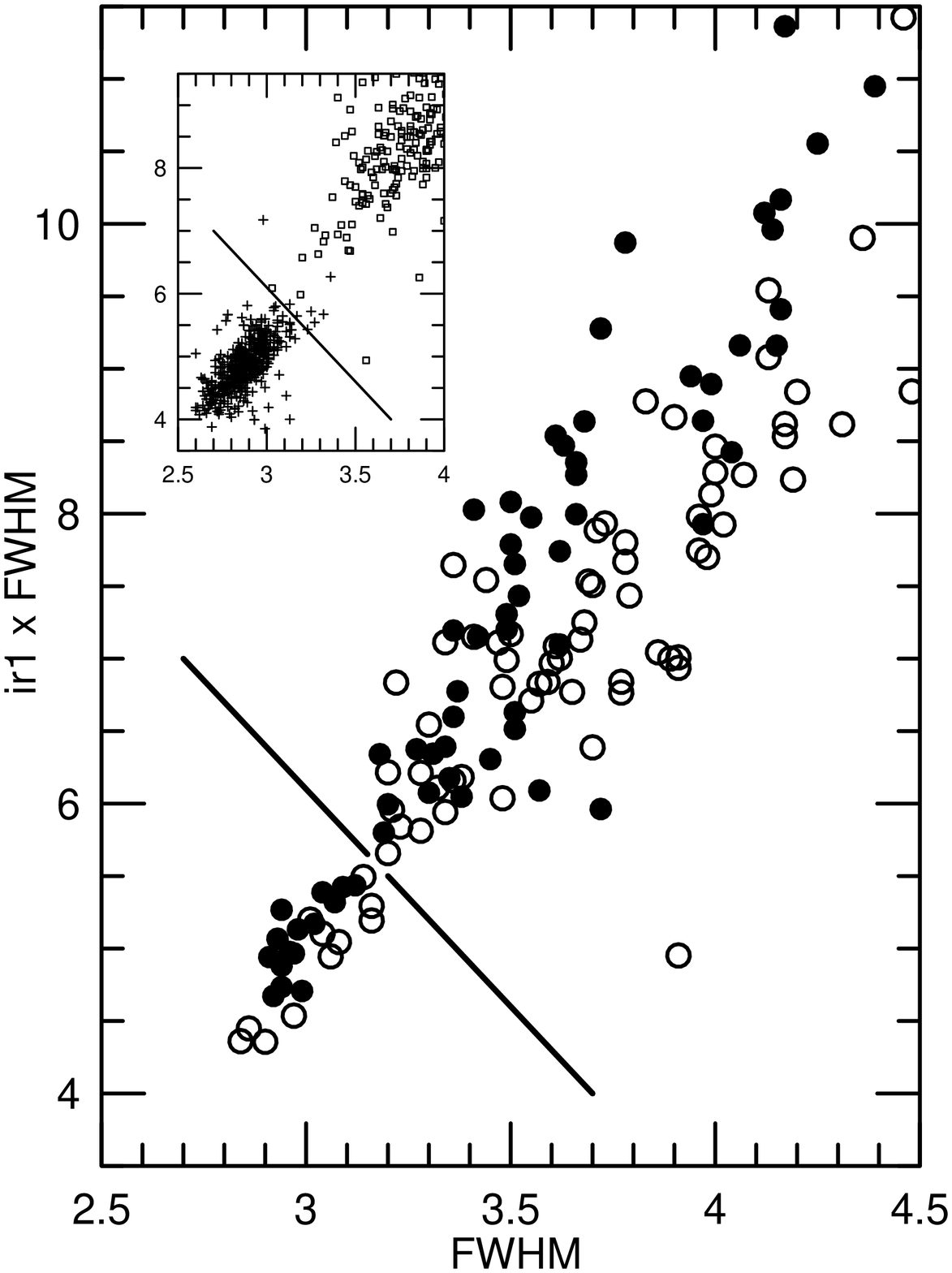]{
 FWHM versus ir1$\times $FWHM diagram used to separate stars and galaxies
 in the bright survey.
 FWHM and ir1 are in the pixel units,
 and the FWHM of PSF was about 3.0 pixels.
 The filled circles represent the objects of $16<K<17$
 and the open circles for those of $17<K<17.5$.
 The objects at the lower-left side of the boundary (bold line)
 are compact and classified as stars,
 while the objects at the upper-right side of the boundary
 extending to outside this figure are diffuse and classified as galaxies.
 The inset shows the plots for the simulated stellar profiles
 and galaxy profiles of $17<K<17.5$ 
 which were used for the completeness and photometry
 simulations in the bright survey.
 The pluses represent the simulated stellar profiles
 and the open squares represent the simulated galaxy profiles.
 The simulated stellar profiles are well confined
 in the lower-left side of the boundary.
 \label{fig4}}

\figcaption[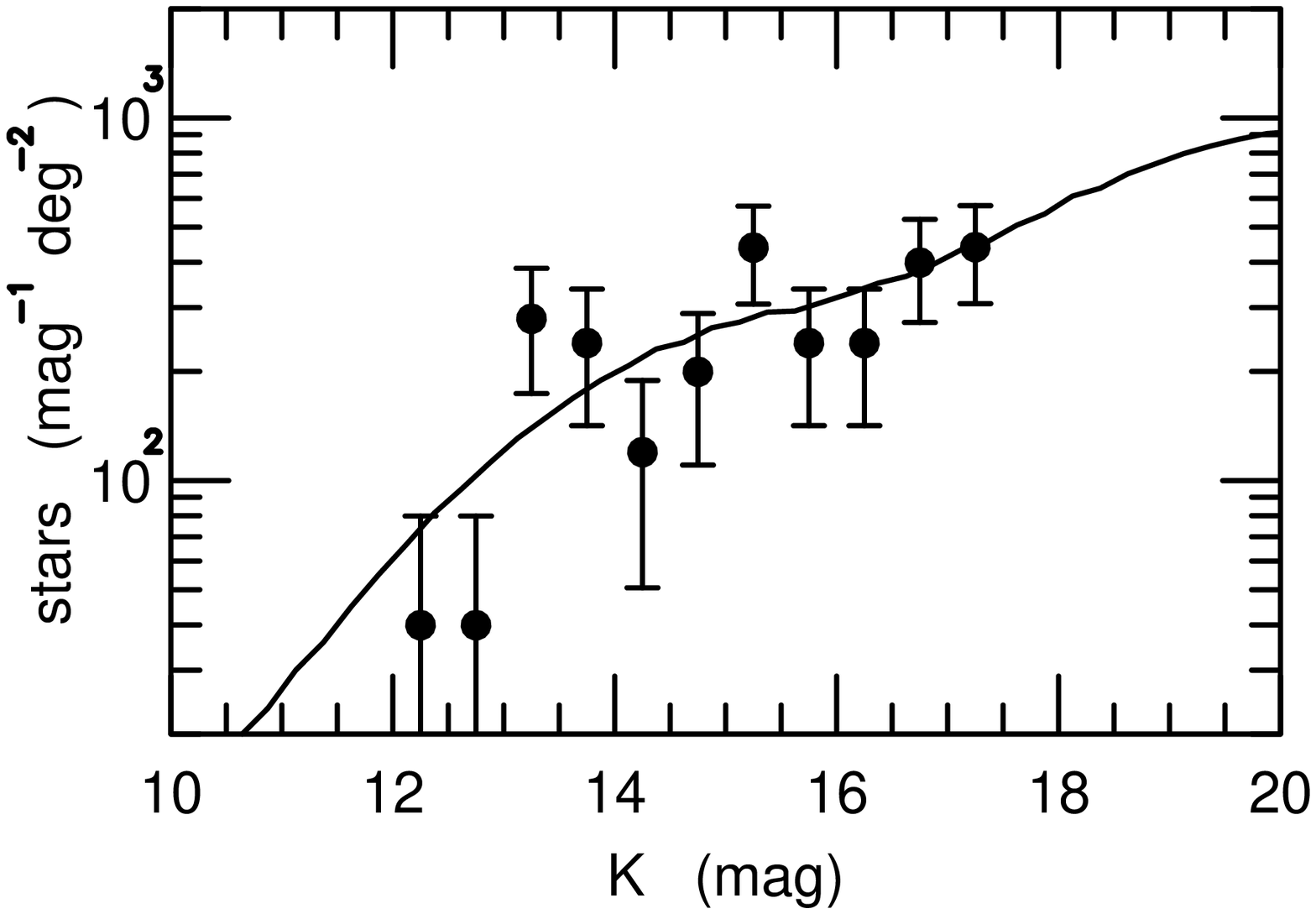]{
 The $K$-band star counts per magnitude per degree${}^{2}$ in the SGP region.
 The filled circles represent the data obtained from the bright survey,
 and the solid line for the fitted model of the SKY version 4 (Cohen 1995).
 \label{fig5}}

\figcaption[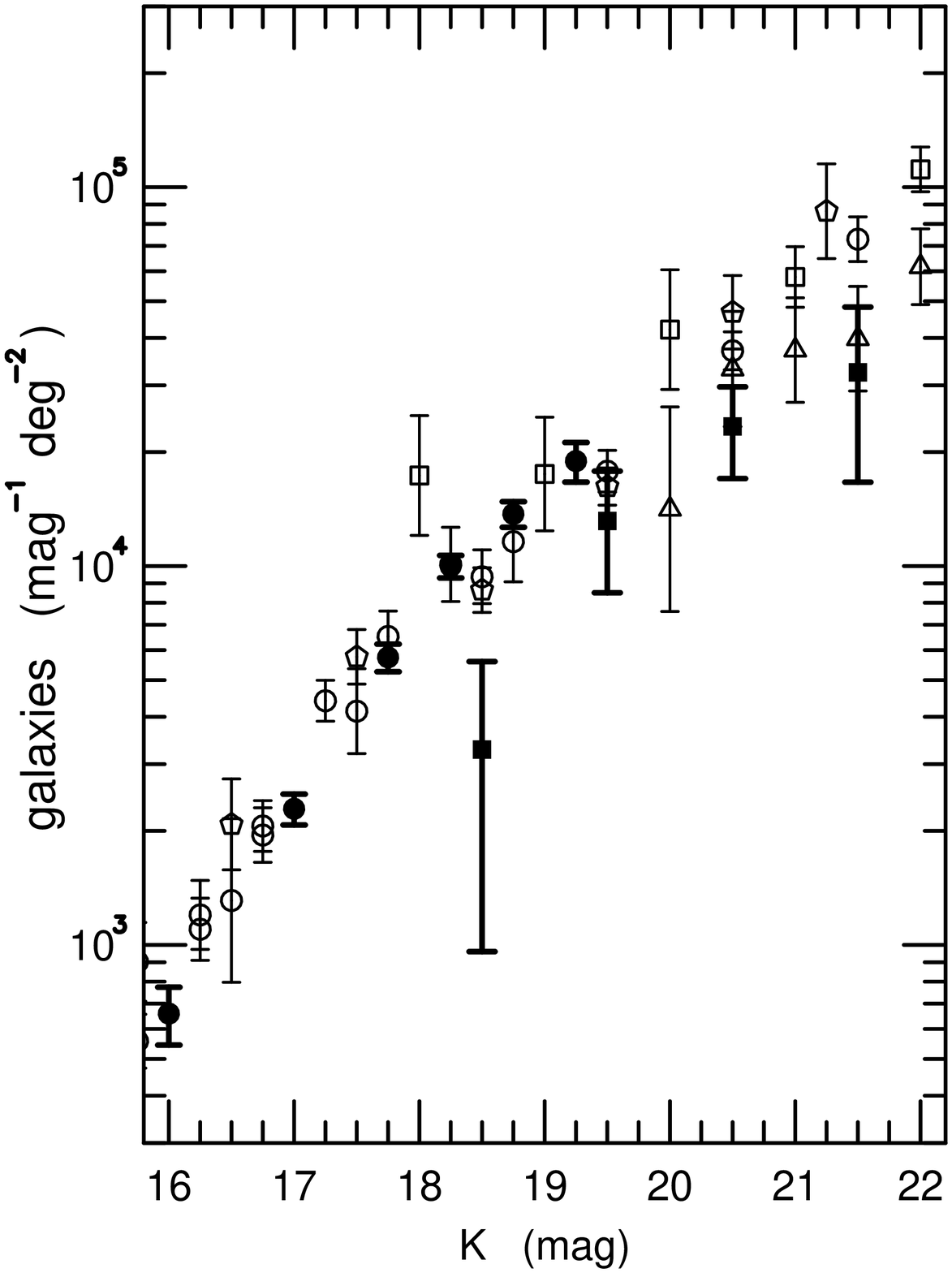]{
 The $K$-band galaxy counts per magnitude per degree${}^{2}$.
 The filled circles represent the data obtained from the bright survey,
 and the filled square represent those from the faint survey.
 We note that the faintest points of each surveys become unreliable
 because their completenesses were so small, $\lesssim $ 50\%,
 and the large corrections were needed.
 The open circles represent the data of the HMWS, HMDS, HDS compiled by
 Gardner et al. (1993) and Cowie et al. (1994),
 the open pentagons for McLeod et al. (1995),
 the open triangles for Djorgovski et al. (1995),
 and the open boxes for Moustakas et al. (1997).
 \label{fig6}}

\clearpage
\begin{deluxetable}{crcrr}
\small
\tablewidth{0pt}
\tablecaption{The $K$-band star count \label{tab05}}
\tablehead{
 \colhead{$K$} &
 \colhead{Raw N\tablenotemark{a}}  & \colhead{Completeness\tablenotemark{b}} &
 \colhead{n\tablenotemark{c}}  & \colhead{Error\tablenotemark{c}}
}
\startdata
 12.0--12.5 &  1 & 0.998 &  39.9    &  39.9\nl
 12.5--13.0 &  1 & 0.998 &  39.9    &  39.9\nl
 13.0--13.5 &  7 & 0.998 & 279.\phd & 106.\phd\nl
 13.5--14.0 &  6 & 0.998 & 240.\phd &  97.8\nl
 14.0--14.5 &  3 & 0.998 & 120.\phd &  69.1\nl
 14.5--15.0 &  5 & 0.998 & 200.\phd &  89.3\nl
 15.0--15.5 & 11 & 0.998 & 439.\phd & 132.\phd\nl
 15.5--16.0 &  6 & 0.998 & 240.\phd &  97.8\nl
 16.0--16.5 &  6 & 0.998 & 240.\phd &  97.8\nl
 16.5--17.0 & 10 & 0.998 & 399.\phd & 126.\phd\nl
 17.0--17.5 & 11 & 0.995 & 440.\phd & 133.\phd\nl
\enddata
\tablenotetext{a}{
 Raw counts of detected stars in the specified magnitude range.}
\tablenotetext{b}{
 The average of completeness for $15\le K\le 17$ was presented at $K\le 17$.}
\tablenotetext{c}{
 Corrected star counts and the errors per magnitude per degree${}^{2}$.}
\end{deluxetable}

\begin{deluxetable}{ccrcrr}
\small
\tablewidth{0pt}
\tablecaption{The $K$-band galaxy count \label{tab1}}
\tablehead{
 \colhead{Survey} & \colhead{$K$} &
 \colhead{Raw N\tablenotemark{a}}  & \colhead{Completeness\tablenotemark{b}} &
 \colhead{n\tablenotemark{c}}  & \colhead{Error\tablenotemark{c}}
}
\startdata
The bright survey
 & 12.5--13.5 &   1\phn \phn & 0.998 &    20.0    &   20.0\nl
 & 13.5--14.5 &   2\phn \phn & 0.998 &    39.9    &   28.2\nl
 & 14.5--15.5 &   8\phn \phn & 0.998 &   160.\phd &   56.5\nl
 & 15.5--16.5 &  33\phn \phn & 0.998 &   659.\phd &  115.\phd\nl
 & 16.5--17.5 & 114.    \phd & 0.996 &  2290.\phd &  214.\phd\nl
 & 17.5--18.0 & 143.    \phd & 0.989 &  5740.\phd &  481.\phd\nl
 & 18.0--18.5 & 246.    \phd & 0.982 &  9980.\phd &  682.\phd\nl
 & 18.5--19.0 & 283.    \phd & 0.821 & 13700.\phd & 1070.\phd\nl
 & 19.0--19.5 & 178.    \phd & 0.375 & 18900.\phd & 2270.\phd\nl
\nl
The faint survey
 & 18.0--19.0 &   2\phd \phn & 0.994 &  3280.\phd &  2320.\phd\nl
 & 19.0--20.0 &   8\phd \phn & 0.991 & 13200.\phd &  4650.\phd\nl
 & 20.0--21.0 &  13.8        & 0.964 & 23400.\phd &  6350.\phd\nl
 & 21.0--22.0 &  10.5        & 0.527 & 32500.\phd & 15800.\phd\nl
\enddata
\tablenotetext{a}{
 Raw counts of detected galaxies in the specified magnitude range.}
\tablenotetext{b}{
 The average of completeness for $15\le K\le 17$ was presented at $K\le 17$.}
\tablenotetext{c}{
 Corrected galaxy counts and the errors per magnitude per degree${}^{2}$.}
\end{deluxetable}

\end{document}